\documentclass{ws}
\def\0{\over } \def\2{{1\over2}} \def\4{{1\over4}}
\def\5{\hat } \def\6{\partial }

   \def\d{\delta }
\def\e{\epsilon }

 \def\o{\omega }

\def\({\left(} \def\){\right)} \def\<{\langle } \def\>{\rangle }

\newcommand{\nn}{\nonumber\\ }
\newcommand \beq{\begin{eqnarray}}
\newcommand \eeq{\end{eqnarray}}

\def\Im{{\,\rm Im\,}}
\def\Re{{\,\rm Re\,}}
\def\tr{{\,\rm tr\,}}

\def\rightnote{{\footnotesize\sl $^\ast$to appear in the Proceedings of 
SEWM2000, Marseille, June 14--17, 2000} \hfill}

\begin{document}

\title{Improved Resummations for the Thermodynamics of the Quark-Gluon
  Plasma$^\ast$
}

\author{A. Rebhan}

\address{Institut f\"ur Theoretische Physik,
         Technische Universit\"at Wien,\\
         Wiedner Hauptstra\ss e 8-10/136,
         A-1040 Vienna, Austria}

\maketitle

\abstracts{
Two recent attempts for overcoming the poor convergence of
the perturbation expansion of the thermodynamic potentials
of QCD are discussed: an HTL-adaption of ``screened perturbation
theory'' and approximately self-consistent HTL resummations
in the two-loop entropy.}


At leading order, perturbation theory in the deconfined phase of QCD
gives a reasonable estimate of the interaction pressure for
temperatures a few times the critical one. But as soon as
the beautiful machinery of resummed thermal perturbation theory
comes into its own, its poor convergence properties
seem to forbid its exploitation except at ridiculously
high temperatures (or densities).\cite{Arnold:1995eb,Braaten:1996ju}
This breakdown becomes apparent already at order $g^3$, which is
entirely produced by the collective phenomenon of Debye screening
(somewhat misleadingly dubbed ``plasmon effect''), and
already occurs in the simplest models such as scalar $\phi^4$ theory
for rather small coupling.\cite{Drummond:1997cw}

At least in scalar theory, it has been shown that this impasse
can be breached by Pad\'e resummation\cite{Drummond:1997cw}
and, more promisingly, by judiciously optimized 
perturbation theory such as ``screened perturbation theory'' 
(SPT)\cite{Karsch:1997gj,Andersen:2000yj}.

In SPT a coupling expansion is performed only with respect to
couplings in explicit interactions, while any coupling constants
buried in thermal (quasiparticle) masses are not expanded out,
leading to nonpolynomial, {\it i.e.} nonperturbative, expressions
in $g$.
This has recently been adapted for
QCD under the 
trademark
``HTL perturbation theory''\cite{Andersen:1999fw}.
There, in place of a simple mass term,
the 
hard-thermal-loop\cite{Braaten:1990mz} (HTL) 
effective action is added, and subtracted
again as a formally higher-order counterterm, from the ordinary
action. 

This approach differs from standard (HTL-)resummed perturbation 
theory\cite{Braaten:1990mz}
in that resummed quantities are not only used in the soft momentum
regime, but throughout. However, there is a price to be paid. At any
{\em finite} loop order,
the UV structure of the theory is modified---new 
(eventually temperature-dependent) divergences
occur and must be subtracted, introducing a new source of
renormalization scheme dependence. 

An alternative approach for a more
extensive resummation of the physics of HTL's
has been worked out by J.-P.~Blaizot, E.~Iancu, and 
myself\cite{Blaizot:1999ip,Blaizot:1999ap,Blaizot:2000fc},
which is based on a self-consistent (``$\Phi$-derivable'')
two-loop approximation to the thermodynamic potentials.
A central observation regarding the latter is that
the {\em entropy} 
has a remarkably simple form,
\be\label{Ssc}
{\cal S}=-\tr\int\!\!{d^4k\0(2\pi)^4}{\6n\0\6T} \Im \log D^{-1} 
+\tr\int\!\!{d^4k\0(2\pi)^4}{\6n\0\6T} \Im\Pi \Re D
\ee
up to terms that are of loop-order 3 or higher,
provided $D$ and $\Pi$ are the self-consistent one-loop propagator and
self-energy. Thus, any explicit two-loop interaction
contribution to the entropy has been absorbed by the spectral
properties of quasiparticles.
Remarkably, this holds true for fermionic\cite{Vanderheyden:1998ph}
as well as gluonic\cite{Blaizot:1999ip,Blaizot:2000fc} 
interactions.

\def\rightnote{}

Now, except for simple scalar models, such a self-consistent
calculation is usually prohibitively difficult. In gauge
theories it is moreover of questionable value because it
is gauge-fixing dependent. However, these gauge dependences
occur at an order which is beyond the (perturbative) accuracy of the
above 2-loop approximation. If only the relevant 
leading and next-to-leading
order contributions to the self-energies are considered, gauge invariance
remains intact. We have therefore proposed {\em approximately
self-consistent} (ASC) resummations based on
Eq.~(\ref{Ssc}) with, in a first approximation,
the HTL 
self-energies, and, in a
next-to-leading approximation (NLA), ones that are
augmented by contributions given by NLO HTL perturbation
theory for hard quasiparticles.

Employing HTL propagators, one obtains an expression, 
${\cal S}^{\rm HTL}$, 
which
is no more complicated than the HTL-resummed {\em one}-loop 
pressure of Ref.%
\citelow{Andersen:1999fw} (in one respect it
is even simpler as it is manifestly UV finite and
does not need artificial subtractions\footnote{As a matter of fact,
the evaluation of the HTL pressure in 
Ref.\citelow{Andersen:1999fw} has recently been
found\cite{Blaizot:2000fc} to suffer from an incomplete dimensional
regularization that led to a larger-than-necessary over-inclusion
of the leading-order interaction term.}). And in contrast to the
latter, when expanded to order $g^2$, it contains the correct
leading-order interaction term, given by 
\be
\label{S2}
{\cal S}^{(2)} =-2\pi N_g\int\!\!{d^4k\0(2\pi)^4}{\6n\0\6T} 
\e(\o)\d(\o^2-k^2)
\Re\Pi_T^{\rm HTL}(\o,k)
=-N_g{m_\infty^2T\06}
\ee
(in pure-glue QCD). Remarkably, this is directly
related to the asymptotic thermal mass $m_\infty^2=Ng^2T^2/6$
of hard transverse gluons. 

On the other hand, ${\cal S}^{\rm HTL}$ contains only part of
the plasmon effect $\sim g^3$; the main contribution $\sim g^3$
comes, rather surprisingly, from 
corrections to the dispersion laws of {\em hard} quasiparticles,
determined by
$\delta \Pi_T^{\rm HTL}$ as evaluated
by standard HTL perturbation theory\cite{Blaizot:2000fc}. 
Both, $\Pi_T^{\rm HTL}$ and
$\delta \Pi_T^{\rm HTL}$ turn out to be needed only for
approximately light-like momenta\footnote{The 2-loop entropy assumes its
simple form of Eq.~(\ref{Ssc}) 
only after the sum over Matsubara frequencies is
carried out and an inherently real-time formula is obtained.}, 
which is gratifying as this
is the only region where the HTL's remain accurate
for hard momenta.

\begin{figure}[t]
\epsfxsize=6.25cm
\centerline{\epsfbox[76 220 540 550]{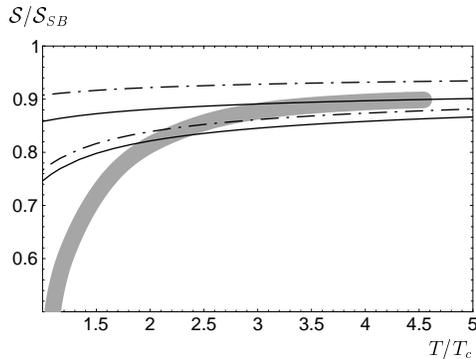}} 
\caption{Pure-glue SU(3) gauge theory:
Comparison of the HTL entropy (full lines) and the NLA estimates
(dash-dotted lines) for
$\overline{\hbox{MS}}$ renormalization scale $\bar\mu=\pi T\ldots
4\pi T$ with the lattice result of Ref.\protect\citelow{Boyd:1996bx}
(dark-gray band).
\label{fig:Sg}}
\end{figure}

In Fig.~\ref{fig:Sg}, ${\cal S}^{\rm HTL}$ has been evaluated
with $g(\bar\mu)$ and $\bar\mu=\pi T\ldots 4\pi T$ and is
found to compare favorably with available lattice 
data\cite{Boyd:1996bx}. 
Also included are estimates for our proposed next-to-leading
approximation (NLA) which corrects the asymptotic thermal mass $\delta
m_\infty$ by the (averaged) NLO contribution as given by standard
HTL perturbation theory, and incorporated through an
approximate gap equation\cite{Blaizot:2000fc} (cf. Appendix). 

This approach has also been applied successfully to QCD with
fermions at zero and non-zero chemical potentials.
Further elaborations and refinements are work in progress.

\begin{figure}[t]
\centerline{\epsfxsize=5.75cm\epsfbox[80 220 540 550]{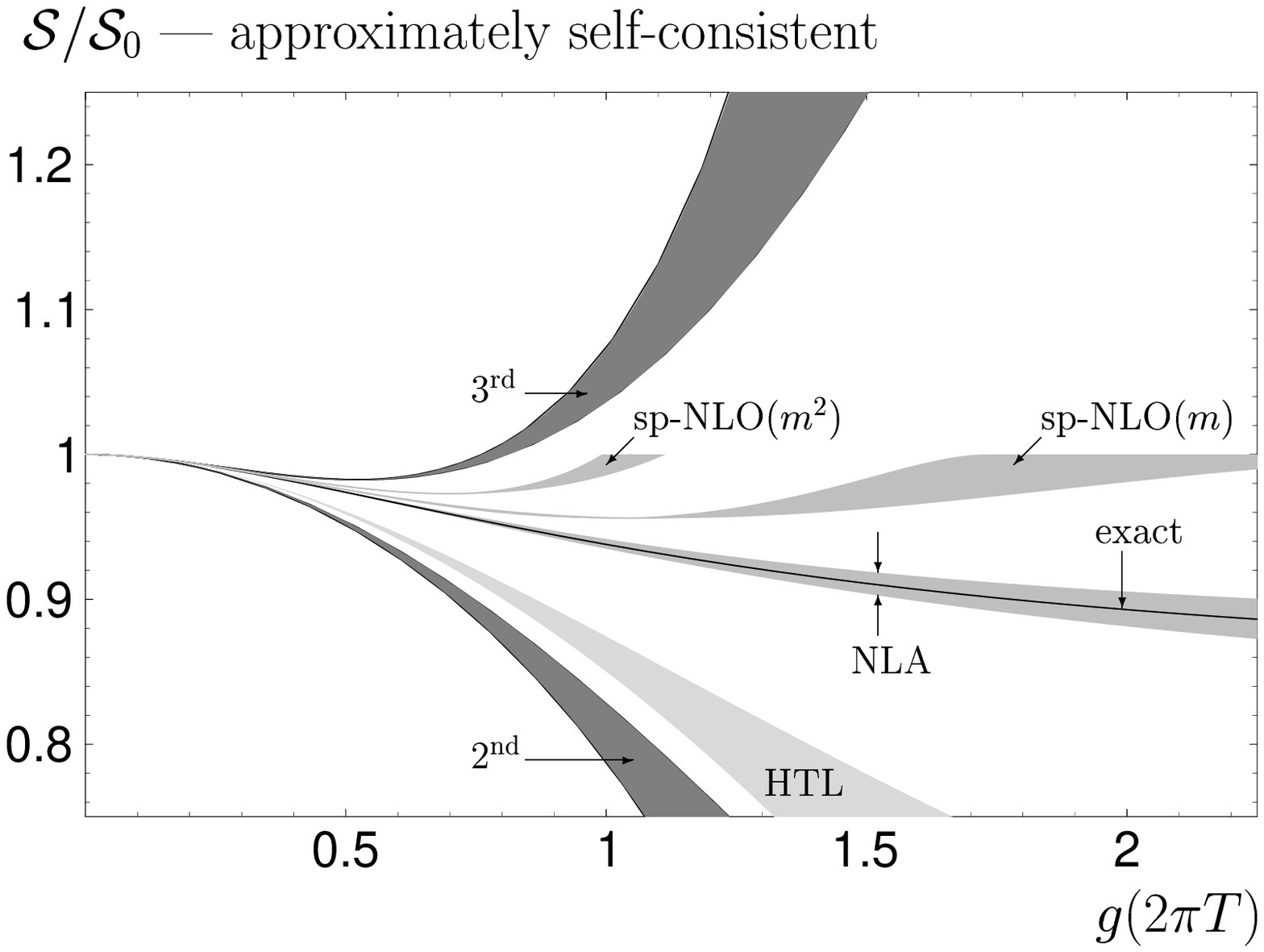}
\epsfxsize=5.75cm\epsfbox[80 220 540 550]{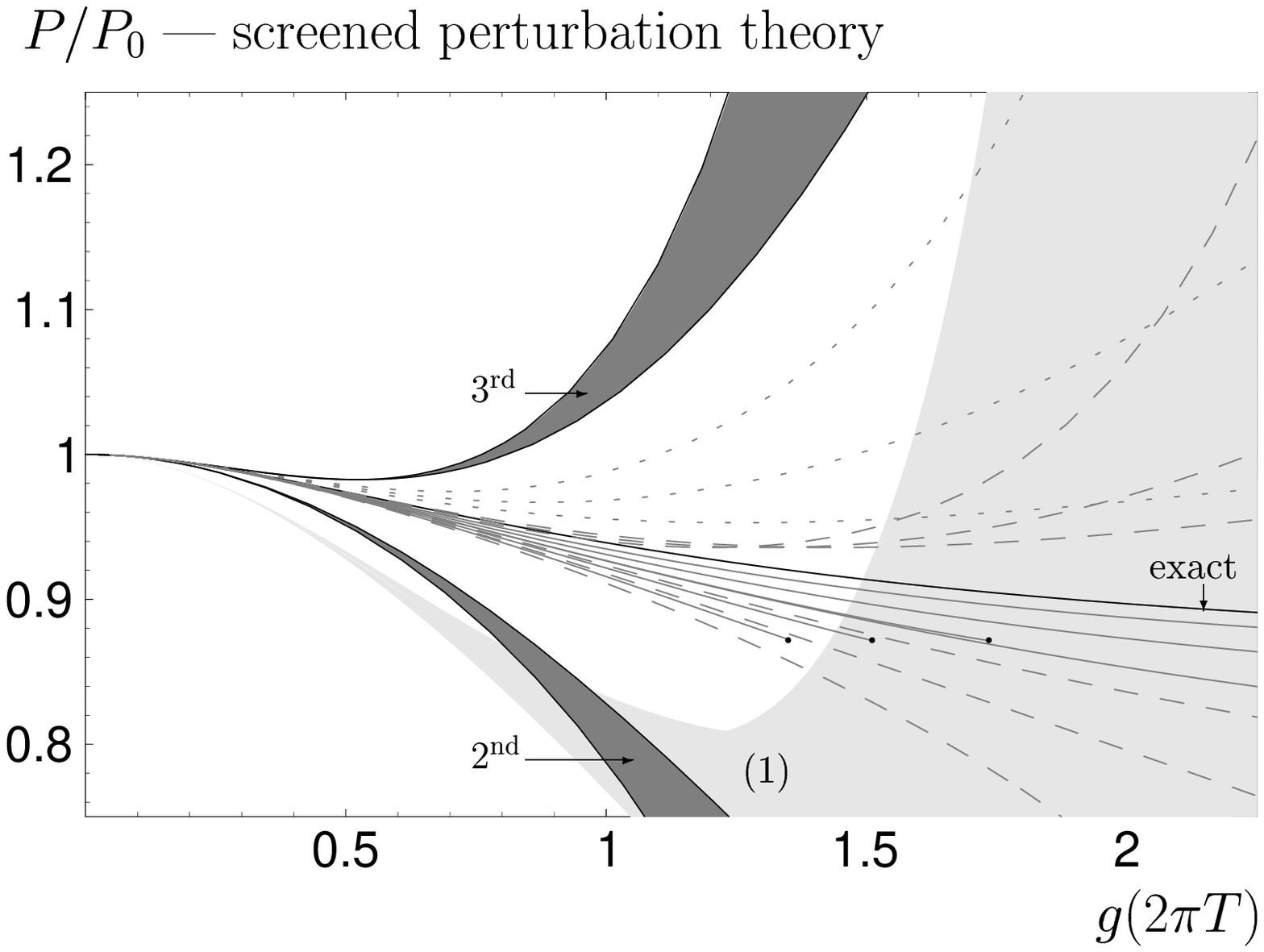}
} 
\vspace{-2mm}
\centerline{\footnotesize\hfil (a) \hfil\hfil\hfil\hfil (b) \hfil}
\vspace{-1mm}
\caption{Large-$N$ scalar O($N$)-model:
(a) Comparison of (2nd- and 3rd-order)
perturbative and HTL-improved approximations to
the entropy
. The shaded areas
denote the variation under changes of the renormalization scale
from $\bar\mu'=\pi T$ to $4\pi T$. The band marked ``HTL'' refers
to using the leading-order (HTL) mass in the 2-loop $\Phi$-derivable
entropy, ``NLA'' to using the approximately self-consistent
mass $m^2=g^2T^2-3g^2Tm/\pi$. Also given are the corresponding
results for a naive strictly perturbative NLO (sp-NLO) mass when defined
through $m^2=g^2T^2-3g^3T^2/\pi$ or $m=gT-3g^2T/2\pi$, respectively.\hfil\break
(b) The analogous comparison for the HTL-resummed one-loop pressure
and the two-loop pressure in screened perturbation theory. The
light-gray
area marked (1) corresponds to the HTL-resummed one-loop pressure
with in addition $\bar\mu_3$ varied from $\2m$ to $2m$.
Full
dark-gray lines refer to the ``minimally subtracted'' 2-loop pressure
in SPT, Eq.~(\ref{P12spt}), where $m$ is chosen by extremalization.
Displayed are the results
for $\bar\mu_3=2m$ (upper three curves corresponding
to $\bar\mu/(2\pi T)=\2,1,2$) and $\bar\mu_3=\2m$ (lower three curves,
which have finite end-points beyond which there is no solution
to the extremalization condition). With $\bar\mu_3=\bar\mu$, the
result coincides with the exact one. The a priori equally plausible
prescription of subtracting $P(0)$ instead, Eq.~(\ref{P12sptth}), 
together with
$\bar\mu_3=\2m$ or $2m$ leads to the various dashed lines,
the choice $\bar\mu_3=\bar\mu$ to the dotted ones.
\label{fig:sc}}
\vspace{-3mm}
\end{figure}

\section*{Appendix}

In Fig.~\ref{fig:sc}a, our approximately self-consistent entropy
is considered for the ``solvable''
toy model of massless O($N\to\infty$) scalar field theory
and compared to the results of screened perturbation theory
at one- and two-loop order.
In this model the
unrenormalized Lagrangian is
$
{\cal L}(x)=\2(\6 \vec \phi)^2 
-{3\0N+2}g_0^2(\vec\phi^2)^2,
$
to be taken in the limit $N\to\infty$, where the
pressure per scalar degree of freedom
coincides with that of $N=1$ when keeping
only 
``super-daisies'' or
``foam'' diagrams\cite{Drummond:1997cw}. As is well known, this leads to
\be\label{PlargeN}
P(T)-P(0)={J}_T(m)+\2 m^2 I_T(m)+{1\0128\pi^2}m^4
\ee
with the thermal mass $m$ given by the solution of the ``gap equation''
\be\label{largeNgap}
m^2=4!g^2(\bar\mu) [I_T(m)+I_0^f(m,\bar\mu)]
=g^2T^2-{3\0\pi}g^3T^2+\ldots
\ee
where $g(\bar\mu)$ has been minimally renormalized and where
we have introduced\goodbreak
\bea
I_0(m)  
&=& -{m^2\032\pi^2}\({2\0\e}+[\log{\bar\mu^2\0m^2}+1]\)
\equiv I_0^{\rm div}(m)+I_0^f(m,\bar \mu)
\\
I_T(m) &=& \int {d^3k\0(2\pi)^3}
        {n(\varepsilon_k)\02\varepsilon_k}, \qquad
{J}_T(m)
=-T\int\frac{{d}^3k}{(2\pi)^3}\,
\log(1-{\rm e}^{-\varepsilon_k/T})
\eea
with $\varepsilon_k=\sqrt{k^2+m^2}$.


In the present context, 
SPT
amounts to 
replacing
$
{\cal L}\to {\cal L}-\2m_0^2\vec \phi^2+\d\2 m_0^2\vec \phi^2
$
where $\d$ is treated as a one-loop quantity prior to putting $\d=1$,
and $m$ is in the end
some approximation to the thermal mass, e.g. as given by
some (approximate) gap equation\cite{Karsch:1997gj,Andersen:2000yj} 
or by the HTL value\cite{Andersen:1999fw} $m=gT$.

Now this introduces new UV divergences, requiring also
a mass counterterm, which however must be subtracted again
in the $\d$-counterterms, for the original theory is massless and
does not have mass counter\-terms in dimensional regularization.
In our simple model, a renormalized mass $m$ can be introduced
by
$
m_0^2=m^2-4!g_0^2 I^{\rm div}_0(m)
$.
The divergences in the two-loop pressure are then formally
$T$-independent (before $m$ gets identified with
some thermal mass), and their minimal subtraction yields
\bea\label{P12spt}
P^{(1)+(2)}_{\rm SPT,min.}(T)&=&
{J}_T(m)-\2 m^2 I_0^f(m,\bar\mu_3)+{m^4 \0 128\pi^2}\nn
+m^2 [I_T(m)&+&I_0^f(m,\bar\mu_3)]-12g^2\{I_T(m)+[I_0^f(m,\bar\mu_3)]\}^2,
\eea
where the first three terms represent the one-loop contribution.
Here $\bar\mu_3$ is the renormalization scale associated with
the additional divergences of SPT.\cite{Andersen:1999fw}

Alternatively, one could, with equal plausibility, 
define a finite pressure by considering
$P(T)-P(0)$. This explicitly thermal part reads
\be\label{P12sptth}
P^{(1)+(2)}_{\rm SPT,th.}(T)=
{J}_T(m)
+m^2 I_T(m)
-12g^2\{I_T(m)^2+2I_T(m) I_0^f(m,\bar\mu_3)\}.
\ee

In Fig.~\ref{fig:sc}b, the (HTL)-resummed 1-loop pressure and the
2-loop pressure of SPT with $m$ fixed by extremalization,
$\6P/\6m=0$, are evaluated for various subtraction schemes.
It turns out that SPT works well only beginning at 2-loop order 
and only in version (\ref{P12spt}), provided $\6P/\6m=0$
has solutions.

In the ASC approach, already the
HTL approximation is a significant improvement over standard
perturbation theory. The NLA works extremely well provided
the NLO corrections to the thermal mass are included by
the ASC gap equation $m^2=g^2T^2-3g^2Tm/\pi$.

\end{document}